\documentclass[11pt,twoside]{article}
\usepackage{asp2004}
\usepackage{psfig}
\usepackage{epsf}
\usepackage{graphics}
\usepackage{lscape}
\def\edcomment#1{\iffalse\marginpar{\raggedright\sl#1\/}\else\relax\fi}
\reversemarginpar

\begin{document}
\title{Catalog of Nearby Galaxies and the Local Cosmic Web}
\author{Igor D. Karachentsev}
\affil{Special Astrophysical Observatory, Russian Academy
	  of Sciences, N.\ Arkhyz, KChR, 369167, Russia}
\author{Valentina E. Karachentseva}
\affil{Astronomical Observatory of Kiev University, Kiev, Ukraine}
\author{Walter K. Huchtmeier}
\affil{
Max-Planck-Institut f\"{u}r Radioastronomie, Auf dem H\"{u}gel 69, D-53121 Bonn,
 Germany}
\author{Dmitry I. Makarov}
\affil{Special Astrophysical Observatory, Russian Academy
of Sciences, N. Arkhyz, KChR, 369167, Russia}

\begin{abstract}
  We compiled an all-sky catalog of 451 nearby galaxies, each having
an individual distance estimate $D \la 10$ Mpc or the radial velocity
$V_{LG} < 550$ km s$^{-1}$. The catalog contains data on basic optical and
HI properties of the galaxies: their diameters, absolute
magnitudes, morphological types, optical and
HI surface brightnesses, rotational velocities, indicative
mass-to-luminosity and HI mass-to-luminosity ratios, as well
as a so-called "tidal index", which quantifies the galaxy environment.
We expect the catalog completeness to be $\sim75$\% within 8 Mpc.
About 85\% of the Local volume population are dwarf (dIrr, dIm, dSph)
galaxies with $M_B>-17.0$, which contribute about 4\% to the local
luminosity density, and $\sim(10-16)$\% to the local HI mass density.

We found that the mean local barion density $\Omega_b(<8$ Mpc) = 2.3\%
consists of only a half of the global barion density,
$\Omega_b=(4.7\pm0.6)$\% (Spergel et al. 2003). The mean-square
pairwise difference of radial velocities is about 100 km s$^{-1}$ for spatial
separations within 1 Mpc, increasing to $\sim 300$ km s$^{-1}$ on a scale of
$\sim 3$ Mpc.
\end{abstract}
\thispagestyle{plain}

Numerous tasks of extragalactic astronomy and cosmology require
to have a quite complete and representative collection of galaxies
limited by a fixed distance. In distinction to the galaxy samples limited
by the apparent magnitude, creation
of the distance-limited sample is an exceptionally difficult problem because
of the huge difference of galaxies according to their luminosity, surface
brightness and other global parameters.
Over the last decade, tremendous advance
has been made in searching for nearby ( $D < 10$ Mpc) galaxies owing to
purpose oriented efforts of different observational teams.
The employment of the Hubble Space Telescope
makes it possible to study the stellar population
and  measure accurate distances of galaxies which are situated outside the
Local Group. It is time for summarizing the wealth of
new observatial
data on neighboring galaxies allowing us to establish some basic
properties of the Local Cosmic Web.

\section{Recent observations of the Local Volume galaxies}

The nearest rich cluster of galaxies in Virgo is situated at a distance
of $D_{Virgo} = 17$ Mpc from us. Therefore, the Local Volume (=LV)
of radius $D = 8$ Mpc, undisturbed by high virial motions,
may be considered as quite a representative volume as to its
population as well to the presence of basic structural properties:
groups, voids, etc. The first step towards compiling such a ``fair''
sample was made by Kraan-Korteweg \& Tammann (1979), who published a
``Catalog of nearby galaxies within 10 Mpc''. The catalog contains 179
galaxies satisfying the condition  $V_{LG} < 500$ km s$^{-1}$, where $V_{LG}$
is the galaxy radial velocity corrected for the observer movement with respect
to the Local Group centroid.

Global parameters in this sample were studied by Huchtmeier \&
Richter (1988). Later, Karachentsev (1994) published an updated version
of the LV list, which contains 226 galaxies with $V_{LG} < 500$ km s$^{-1}$.
Following the same condition, Karachentsev, Makarov \& Huchtmeier (1999b)
enlarged the initial LV sample to 303 objects basing on new optical
and HI observations.
Over the past few years, special searches for new nearby dwarf galaxies
have been undertaken basing on the optical sky survey POSS-II/ESO/SERC,
HI and infrared surveys of the Zone of Avoidance, ``blind'' sky surveys in
the 21 cm line, HIPASS and HIJASS. Armandroff et al. (1999) carried out
automated searches for dwarf companions of low surface brightness in a
wide vicinity of the Andromeda galaxy (M~31). Karachentsev \& Karachentseva
with collaborators undertook a visual inspection of all POSS-II/ESO/SERC
plates and found more than 500 nearby dwarf galaxy candidates, mainly of
low surface brightness.
HI observations of these objects (Huchtmeier et al. 2000, 2001)
as well as optical spectral observations (Makarov et al. 2003) revealed
about 100 new galaxies with radial velocities less than 500 km s$^{-1}$.
The nearest
galaxy group around IC 342/Maffei behind the Milky Way was studied
in the HI line and in IR that resulted in finding some new members of the
group (Buta \& McCall, 1999).
In the region of another nearby group in Centaurus, new gas-rich
dwarf galaxies were found by Kilborn et al., 2002 and Staveley-Smith et al.,
1998. As a result of the joint efforts, the local number density of
galaxies has been increased 2 times.

It is obvious that the condition  $V_{LG} < 500$ km s$^{-1}$ for selection of
nearby galaxies seems to be too simplified. Indeed, the kinematic distance
of a galaxy in a group may be in error by several Mpc because of virial
motions. Apart from peculiar motions, the kinematic distance
is also affected by anisotropic expansion of the Local volume.
According to Karachentsev \& Makarov (1996), the local velocity
field on a scale of $\sim 8$ Mpc is characterized by the Hubble
tensor $H_{ij}$, which has the main values of $H_{xx}:H_{yy}:H_{zz}=
(81\pm3):(62\pm3):(48\pm5)$ in km s$^{-1}$Mpc$^{-1}$. The
minor axis of the corresponding ellipsoid is aligned with the polar
axis of the Local Supercluster, and the major axis has an angle of
$(29\pm5)\deg$ with respect to the direction to the Virgo cluster core.
Therefore, the effect of local anisotropy generates a considerable difference
in kinematic distances of galaxies seen in different directions.

Until recently, the majority of very nearby galaxies had no reliable
direct distance estimates. In the 90s many neighboring spiral and irregular
galaxies were resolved into stars for the first time, which allowed
determination of their distances from the luminosity of blue and red
supergiants with an accuracy of $\sim 25$\%. During the last few years, about
150 nearby galaxies have been imaged with the WFPC2 on the Hubble
Space Telescope (HST). Accurate distances to them (with an error $\sim 10$\%)
were measured from the luminosity of the tip of the red giant branch (TRGB).
A compilation of 223 galaxies situated within 5.5 Mpc from us has been
presented by Karachentsev et al. (2003). About half of them have accurate
distance estimates via TRGB or cepheids. The ACS camera that has recently
been installed on the HST is able to measure distances of galaxies
to within 7--8 Mpc in a fast ``snapshot'' mode (1 object per 1 orbit).
This stimulated us to prepare a sample of galaxies with the
expected distances within 7--8 Mpc, being as complete as possible.

\section{Catalog}

Taking into account the presence
of non-Hubble motions, as well as distance measurement errors, we selected
galaxies for our sample on the basis of two simple conditions:
$D<10\;\; {\rm Mpc}$,
if a galaxy has an individual distance estimate, or
$V_{LG}<550 \;\;{\rm km\; s}^{-1}$,
if the galaxy distance has been estimated from its radial velocity alone.
Such an approach permits us to save a maximum number of galaxies in the
sample with true distances  $D<8$ Mpc, although it includes inevitably
a fraction of more distant objects too.

At total, 451 galaxies satisfy the above mentioned condition (Karachentsev
et al. 2004). Distributions of the LV galaxies according to their absolute
magnitudes, linear diameters, and rotational velocities are presented
in Fig. 1. Assuming our sample within $D=2$ Mpc to be complete
to nearly 100\%, we derive an estimate of completeness within 8 Mpc
to be $\sim75$\%. Therefore, new more careful searches for faint
neighboring galaxies may lead to discovery of $\sim100$ new galaxies
within 8 Mpc around us. The absolute magnitude --- the mean surface
brightness relationship for 451 neighboring galaxies is plotted in Fig. 2.
The mean surface brightnesses spreads over a range of 7 magnitudes. Evidently,
the galaxies of very low surface brightness can easily be lost when
situated in the Zone of Avoidance. McGaugh \& Blok
(1997) supposed that the true distribution
of galaxies according to their mean surface brightness extends beyond
$\Sigma_B \sim 27^m/\sq\arcsec$, and, probably, about 80\% of all galaxies
lie for us beyond the present threshold of detection. Fig. 2 shows that the
mean surface brightness decreases from giant galaxies to dwarfs. Such
tendency is expected when giant and dwarf galaxies have approximately one
and the same mean volume density of stars, that corresponds to the
relation $\Sigma_B \sim (M_B)/3$ shown in Fig. 2 by the dased line.

\section{Galaxy distribution within 8 Mpc}

The sky distribution of 451 nearby galaxies is presented in Fig. 3 in
equatorial coordinates. The distribution looks extremely inhomogeneous,
showing two large empty areas: the Local Void (Tully, 1988) in
Hercules-Aquila and the Local Mini-void (Karachentsev et al., 2002)
in the Orion constellation. Spatial distribution of the galaxies is seen in
Fig. 4 in the Supergalactic coordinates. Galaxies with the distances less and
more than 8 Mpc are shown by large and small circles, respectively. One
can distinguish in Fig. 4a ( the Supergalactic plane projection) some
relatively compact groups around the Milky Way, M~31, M~81, IC~342,
Cen~A, M~83, and also the Canes Venatici cloud. Remarkably, in the huge
volume of the Tully void ($\sim100$ Mpc$^3$) there is not any galaxy with
luminosity brighter than $L\sim 2\cdot 10^6\cdot L_{\sun}$. The dwarf
galaxy KK246 ($-12\fm96$)
is situated just at the edge of the Tully void, but not inside it. The Local
Supercluster center in Virgo
is characterized by the Cartesian coordinates: $SGX=-4$ Mpc, $SGY=16$ Mpc,
$SGZ=-1$ Mpc. A small density gradient is seen towards Virgo, but it is
masked by strong density fluctuations caused by voids. The majority of the
groups locate in a thin layer $\mid SGZ \mid  < 0.3 $Mpc of the Supergalactic
plane. However, there are also groups around NGC~6946, M~101, M~96 (Leo-I),
situated at a distance of $\sim3-4$ Mpc from the Local ``sheet''.

\section{Environment effects in the Local Volume}

As it is know, the HI abundance in disk-like galaxies depends on their
environment. Spiral galaxies in the cores of rich clusters demonstrate
significant HI- deficiency with respect to the field galaxies of the
same morphological type. However, observational data on the HI- deficiency
outside rich clusters look rather controversial. To describe the local
mass density around a galaxy ``$i$'', Karachentsev \& Makarov (1999a)
have introduced the so-called ``tidal index'':
$\Theta_i = \max  \{\log(M_k/D_{ik}^3)\} + C,\;\;\;  i = 1, 2...  N$,
where  $M_k$ is the total mass of any neighboring galaxy separated from
the considered galaxy by a space distance $D_{ik}$. For every galaxy ``$i$
we found its ``main disturber'', producing the highest tidal action or a
maximum density enhancement, $\Delta\rho_k\sim M_k/D^3_{ik}$. In order to take
account of all surrounding galaxies (but not only ones with measured $V_m$),
we determined the total mass of every galaxy from its luminosity. Here,
we accept the mean value $M_{25}/L = 3.8 M_{\sun}/L_{\sun}$
for all morphological types, and also adopted that the total mass of each
galaxy is on average 2.5 times its indicative mass, $M_T=2.5 M_{25}$. The
value of the constant $C$  is chosen so that $\Theta=0$ when the Keplerian
cyclic period of the galaxy with respect to its main disturber equals the
cosmic Hubble time, $1/H$. In this sense, galaxies with $\Theta<0$ may be
considered as undisturbed (isolated) objects.
To characterize the HI-deficiency of galaxies in groups, we considered two
parameters: the morphological type and the HI surface brightness,
$\Sigma_{HI}=M_{HI}/A^2_{25}$, having lower dispersion with respect to
$M_{HI}/L$ and $M_{HI}/M_{25}$ . The relations $T$ versus $\Theta$ and
$\Sigma_{HI}$ versus $\Theta$ are plotted on the upper and lower panels of
Fig. 5, respectively. The bulge-dominated galaxies with $(T<4)$ are shown
by large circles. As can be seen, the data on the upper panel are indicative
of the known morphological segregation effect when E, S0, dSph occur usually
in groups, but not in the general field. Among the most isolated galaxies
with $\Theta< -2.0$ there are exceptionally objects of the latest types,
Ò = 7 - 10. For spiral and irregular galaxies themselves, one can see
a slight (30\%) decrease in the HI surface brightness from the most isolated
( $\Theta< 0$) galaxies towards the most disturbed. This observational
fact may be used to choose between two extreme scenarios of galaxy
evolution when a) star formation in a galaxy is driven by its gas consumption,
or b) the structure of the stellar and gaseous components of a galaxy is
defined by the process of recurrent merging of galaxies.

\section{Some cosmologic parameters extracted from the LV sample}

Based on the derived data, we estimated some parameters important for
cosmology, in particular, the mean local luminosity density
within a sphere of radius $D$ in $L_{\sun}$/Mpc$^3$.
Apart from the peak produced by the Local Group, the luminosity density
shows a secondary peak at D = 3.7 Mpc, caused by the M81 and CenA groups, and
then decreases smoothly to the value $3.5\cdot10^8 L_{\sun}$/Mpc$^3$
at $D$ = 8 Mpc.  Subsequent decrease in the luminosity density occurs because
of the fractal nature of the matter distribution, as well as because of
incompleteness of our sample. The global value of $\rho_{lum}$ in
the B-band was estimated on the basis of essentially deeper samples.
Blanton et al. (2003) and Liske et al. (2003) have derived from the
Sloan Digital Sky Survey and the Millennium Galaxy Catalogue the values:
$1.23\cdot10^8 L_{\sun}$/Mpc$^3$ and $1.43\cdot10^8 L_{\sun}$/Mpc$^3$,
respectively. Both the quantities are reduced to the local value of the Hubble
parameter, $H=72$ km s$^{-1}$Mpc$^{-1}$. It should be noted, however, that the
global estimates of $\rho_{lim}$ were derived when internal extinction
in galaxies was not taken into account. Ignoring the internal extinction,
we obtain $\rho_{lum}(<8 Mpc) = 2.5\cdot10^8 L_{\sun}$/Mpc$^3$, then the
mean local luminosity density exceeds 1.7--2.0 times the global density
in spite of the presence of the Tully void and the absence of rich clusters
in the Local volume.

There is a pessimistic opinion that
about 80\% dwarf galaxies are hidden from us because of their extremely
low surface brightness (McGaugh,1996, Impey \& Bothun, 1997). Such a
strong argument can be checked in principle by observations.
Dwarf galaxies having absolute magnitudes fainter than -17.0
with their relative number of 85\% make a minor contribution,
$\sim(2-5)$\%, to the integrated luminosity or
integrated ``indicative'' mass of the neighboring galaxies. However, their
contribution in the HI mass (10-16)\%, as well as in the sky area covered by
galaxies (31\%), turns out to be more significant. The supposed presence
of an about 5 times larger population of ELSB dwarf galaxies would
lead to an unusually high number of absorption lines seen in QSO spectra.

To characterize the average virial galaxy motions on different scales,
the mean-square difference of radial velocities of two galaxies,
$<\Delta V^2_{12}(R_p)>$, as a function of their projected linear distance
is often used. It is supposed that this quantity allows one to estimate
the dark matter density on corresponding scales. The existing estimates
of $<\Delta V^2_{12}(R_p)>$, obtained from different catalogs of galaxies and
surveys, differ essentially from each other but lie in the range from
200 km s$^{-1}$ (Branchini et al. 2001) to
600 km s$^{-1}$ (Zehavi et al. 2002),
which leads to $\Omega_{DM}\sim(0.1-0.3)$ on a scale of $\sim1$ Mpc.
It should be emphasized that the derived quantities were obtained for galaxy
samples restricted by the flux, but not the distance. Based on the data
of our Catalog, we calculated $<\Delta V^2_{12}>$ as a function of space
separation as well as projected separation between the galaxies.
The mean-square difference of radial velocities changes
slightly from 110 km s$^{-1}$ to 90 km s$^{-1}$ on a scale of $R<1$ Mpc,
apparently reflecting the Keplerian motions in tight galaxy pairs.
The increase in $<\Delta V^2_{12}>$ to $\sim 300$ km s$^{-1}$, seen on scales
of 1--4 Mpc, is caused by the increasing role of the systematic Hubble
component at larger mutual galaxy separations.
As one can see, the projection effects distort
essentially the observed behaviour of $<\Delta V^2_{12}(R)>$, which makes
the interpretation of the relation in terms of $\Omega_{DM}$ rather
uncertain.

\acknowledgements{

 This work was supported by RFFI grant 04-02-16115 and
DFG-RFBR grant 02-02-04012. This search has made use of the NASA/IPAC
Extragalactic Database (NED), the Lyon Extragalactic Database (LEDA),
the HI Parkes All Sky Survey (HIPASS), and the Two Micron All-Sky Survey
(2MASS).

\end{document}